\documentclass[prd,aps,onecolumn,a4paper,floatfix,nofootinbib,notitlepage]{revtex4-1}

\usepackage[utf8]{inputenc}
\usepackage{graphicx,psfrag}
\usepackage{mathrsfs}
\usepackage{amsmath,amsfonts,amssymb}
\usepackage{comment}
\usepackage{xcolor}
\usepackage{enumerate}
\usepackage{array}
\usepackage{hyperref}
\usepackage{boldline,multirow}
\usepackage{lineno}
\usepackage{siunitx}
\usepackage{framed}
\hypersetup{
    colorlinks = true,
    linkcolor = {blue},
    citecolor = {blue},
    urlcolor = {blue},
    linkbordercolor = {white},
    }

\begin{document}


\title{Negative absolute temperature attractor in a dense photon gas}

\author{M. \surname{Ferraro}$^{1,\dag}$}
\author{F. \surname{Mangini}$^{2,3,\dag,*}$} 
\author{K. \surname{Stefańska}$^3$}
\author{W.A. \surname{Gemechu}$^3$}
\author{F. \surname{Frezza}$^3$}
\author{V. \surname{Couderc}$^4$}
\author{M. \surname{Gervaziev}$^{5,6}$}
\author{D. \surname{Kharenko}$^{5,6}$}
\author{S. \surname{Babin}$^{5,6}$}
\author{S. \surname{Wabnitz}$^3$}

\affiliation{${}^1$Dipartimento di Fisica, Università della Calabria, via P. Bucci, 87036, Rende, Italy}
\affiliation{${}^2$Department of Engineering, Niccolò Cusano University, via Don Carlo Gnocchi 3, 00166 Rome, Italy}
\affiliation{${}^3$Dipartimento di Ingegneria dell'Informazione, Elettronica e Telecomunicazioni, Sapienza Universit\`a di Roma, via Eudossiana 18, 00184 Rome, Italy}
\affiliation{${}^4$Université de Limoges, XLIM, UMR CNRS 7252, 123 Avenue A. Thomas, 87060 Limoges, France}
\affiliation{${}^5$Department of Physics, Novosibirsk State University, Pirogova 1, 630090 Novosibirsk, Russia}
\affiliation{${}^6$Institute of Automation and Electrometry, SB RAS, 630090 Novosibirsk, Russia}
\affiliation{${}^\dag$ These authors contributed equally}
\affiliation{${}^*$Corresponding Author: fabio.mangini@uniroma1.it}


\date{\today}

\maketitle

\section*{Abstract}


Statistical mechanics permits to connect the macroscopic properties of matter with the laws governing the evolution of its microscopic constituents. Such an approach has been very successful for systems of particles governed by either classical or quantum mechanics. In a classical gas, different thermodynamic laws apply to the weakly or strongly interacting particles of an ideal or real gas, respectively. Here, we demonstrate that a similar situation occurs for a gas of photons, which is contained in a finite-dimensional box such as a multimode waveguide. We use a few-mode system provided by a standard step-index fiber operated below cutoff, which permits to prepare a high-density gas of photons. We show that, owing to the attractive potential energy contribution to the photon energy induced by the nonlinear Kerr effect, the mode population exhibits a spontaneous inversion from the fundamental to the highest-order mode, as the input laser beam power grows larger. This inversion of the mode power distribution leads to a stable attractor for the output beam, and is associated with a progressive increase of the optical temperature until a flip of its sign leads to a new regime of negative absolute temperatures. Our work demonstrates the ability to all-optically control the shape of laser beams, which is a prerequisite for applications in high-power laser sources, nonlinear imaging, and optical communication systems.


\section*{Main}

In quantum electrodynamics, it is well known that high-energy photons such as $\gamma$ rays may interact with each other, e.g., via vacuum fluctuations. On the other hand, in a classical framework, photon-photon interactions are virtually impossible to occur: photons lack inertia because of their massless nature. Yet, even in the absence of quantum effects, it is well known that the nonlinearity of the material polarizability may permit the exchange of energy and momentum among photons at optical frequencies. This picture of nonlinear optics has led to a thermodynamic description of classical light as a gas of photons trapped in a box \cite{wu2019thermodynamic}. Such a box that confines the photon gas may be simply provided by a multimode waveguide. For example, in optical fiber, a light beam can be expressed as a finite superposition of modes. Their number, denoted as $M$, defines the effective volume accessible to the photon gas.

When nonlinearity is sufficiently weak, the photon gas dynamics is governed by the momentum exchange among photons, akin to an ideal (low-density) gas where particles only possess kinetic energy, and interact via their collisions. This analogy allows one to define an optical temperature $T$ and an optical chemical potential $\mu$, which determine the mode power distribution at thermal equilibrium: the Rayleigh-Jeans law — the classical counterpart of the Bose-Einstein distribution \cite{zanaglia2024bridging}. These thermodynamic parameters, associated with the conservation of light momentum and power, respectively, have purely statistical meanings, i.e., they cannot be directly measured, but can only be calculated based on the equilibrium state of the photon gas.
Interestingly, optical temperatures may have both positive and negative absolute values, as it occurs, for example, in many physical systems \cite{BALDOVIN20211}: consider, for example, localized spin systems \cite{PhysRev.81.279,RevModPhys.69.1,PhysRevLett.106.195301}, ultracold atoms in optical lattices  \cite{doi:10.1126/science.1227831}, photonic and nonlinear lattices \cite{PhysRevLett.130.063801,doi:10.1126/science.ade6523,PhysRevLett.134.097102}. In multimode optical fibers, for $T>0$ the fundamental mode is the most populated; whereas for $T<0$ the highest-order $M$-th mode has the highest occupancy. At $T=\pm \infty$, the power is equally distributed among all of the fiber modes, while $T=0$ represents a singularity \cite{PhysRevLett.130.063801}.

A striking example of wave thermalization in the weakly nonlinear regime at positive, near-zero temperatures is provided by the beam self-cleaning effect, observed in highly multimode, weakly nonlinear optical fibers \cite{krupa2017spatial}. Here, the fundamental mode dominates so widely that, as a result of the thermalization process \cite{pourbeyram2022direct,baudin2023rayleigh,ferraro2023spatial}, the beam at the fiber output acquires a bell-shaped profile, with a corresponding relatively high spatial quality. As such, the beam self-cleaning effect has naturally found application in the development of multimode fiber lasers, enabling the scaling-up of power with respect to state-of-the-art laser sources based on singlemode fibers \cite{teugin2020single,wright2017spatiotemporal}. It should be pointed out that, although it works well to reproduce experiments in the weakly nonlinear regime, applying the theory of ideal gases to nonlinear multimode fibers rests on an underlying contradiction. Namely, the observed mode power redistribution is a result of fiber nonlinearity, yet the theory is purely linear. The aim of this work is to solve this puzzle.

As a matter of fact, for a fiber of finite length, the occurrence of beam self-cleaning is only observed within a certain range of input beam powers. A lower power threshold should be reached to ensure thermalization. Moreover, whenever the nonlinear contribution to the beam dynamics becomes relatively strong, the beam quality is progressively degraded \cite{mangini2023high}. From a thermodynamic perspective, such a spoiling of beam self-cleaning occurs whenever the potential energy of the photons forming the gas is no longer negligible, when compared with their kinetic energy. This raises the question of whether, in a regime where the material nonlinearity is sufficiently strong, multimode optical beams could be treated as Van der Waals or real, as opposed to an ideal photon gas. As a matter of fact, a comprehensive theory of dense or Van der Waals photon gases has yet to be developed. However, such a situation has been recently experimentally probed by Kirsch et al.: they demonstrated the occurrence of an effect akin to the Joule–Thomson expansion in a waveguide array optical system \cite{kirsch2025observation}. 

Here, we theoretically and experimentally demonstrate a novel property of dense photon gases: the spontaneous transition from positive to negative absolute temperatures. To emphasize effects due to the non-ideal nature of a photon gas, the most favorable setup is provided by a few-mode optical fiber. Contrary to the highly multimode systems that have been studied so far, in a few-mode optical waveguide, the contribution of the potential energy can be greatly enhanced, thanks to the reduced volume that is accessible to the dense photon gas. 
In this way, we could uncover a novel phenomenon, with no counterpart in classical thermodynamics: by increasing the input beam power, one observes a flip of the sign of its absolute temperature, which is testified by an inversion of the mode power distribution. In other words, at low powers, in the ideal gas regime, the beam is dominated by the fundamental mode of the fiber. Whereas at high powers, in the real gas regime, the beam is dominated by the highest-order mode. This inversion of temperature and mode distribution is due to the non-negligible contribution of photon-photon interactions in the total potential energy of the multimode optical system.  

\begin{figure}[!ht]
    \centering   \includegraphics[width=0.8\linewidth]{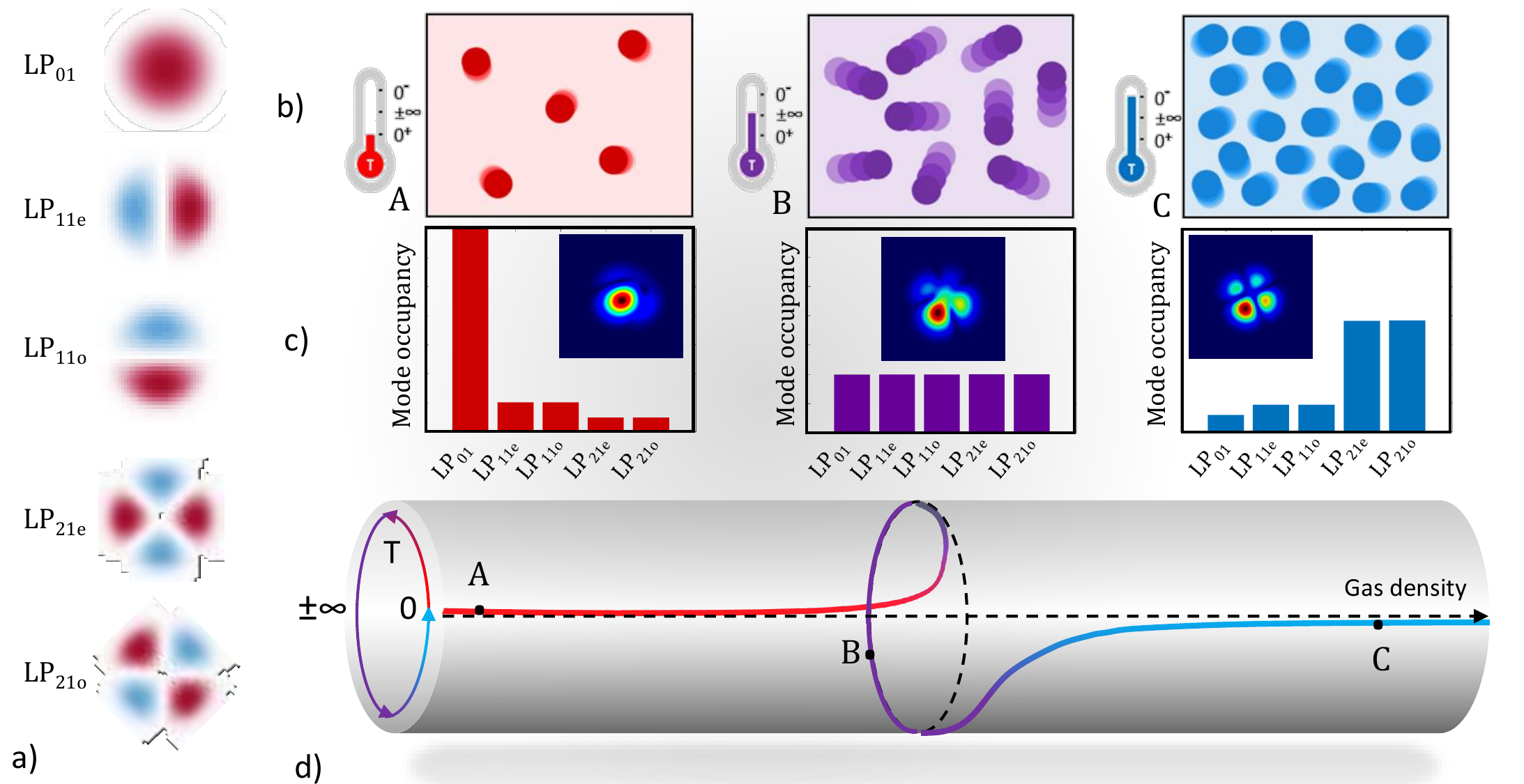}
    \caption{Spatial beam freezing at a glance. (a) Depiction of the first $M$ = 5 guided modes of a step-index fiber sorted by their propagation constant (from top to bottom). (b) Photon gas representation: low density, i.e., ideal, photon gas at positive and close to zero temperature (A); power equipartition among the modes corresponding to $T=\pm \infty$ (B); modal population inversion occurring when the Van der Waals photon gas density is too high and the temperature approaches the negative zero value (C). 
    Configuration B is reached whenever the photon gas density reaches a critical value. The gas is no longer ideal, and the beam may have a speckled profile. Meanwhile in C, the highest order mode is the most populated, and the beam resembles it.
    (c) Mode occupancy distribution and beam intensity profile (inset) corresponding to the states (A-C) in (b). (d) Representation of the temperature evolution as a function of the gas density. Here, we use a cylindrical representation of the temperature values so that the points $T = + \infty$ and $T = - \infty$ coincide. 
    }
    \label{fig:toy}
\end{figure}

To better understand the physical mechanisms behind this effect, consider the following picture. The optical power plays the role of the number of particles forming the gas. Hence, injecting more power into a few-mode optical fiber is equivalent to squeezing the photons into a small finite volume. As the gas density grows, the temperature, which we suppose to be positive at low powers, increases until it eventually reaches and overcomes infinity, up to the point where the photons become packed into a frozen-like ordered state (see Fig. \ref{fig:toy}). This corresponds to a negative and close to zero temperature, i.e., an opposite regime to that which is associated to the beam self-cleaning regime. For this reason, we name the observed effect "spatial beam freezing".

\section*{Conservation laws}
Let us consider the case of a standard optical fiber that supports $M$ modes. Each mode may be characterized in terms of its propagation constant ($\beta$), which depends on the refractive index profile of the fiber. Modes may be sorted according to the magnitude of $\beta$, i.e.,  $\{ \beta \} = \{ \beta_1, \beta_2, ... , \beta_M\}$, where $\beta_1 = \max\{\beta\}$ identifies the fiber's fundamental mode, whereas $\beta_M = \min\{\beta\}$ is the propagation constant of the highest-order mode. The dependence of the propagation constant on the mode number is shown in Fig. \ref{fig:theo}a.
The power carried by the $i$-th mode is $|c_i|^2$; the total power reads as
\begin{equation}
    \mathcal{P} = \sum_{i=1}^M |c_i|^2.
    \label{eq:power}
\end{equation}
is a conserved quantity, as long as the fiber is short enough, so that any linear propagation losses can be neglected. A second conserved quantity is the momentum of light, or Hamiltonian ($H$): in the thermodynamic framework, it is defined as the opposite of the internal energy $U$, i.e., $H=-U$. The internal energy consists of two components: $U=U_L + U_{NL}$. The former, which consists of the kinetic part of the photon gas energy, is associated with linear beam propagation and reads
\begin{equation}
    U_L = - \sum_{i=1}^M \beta_i |c_i|^2,
    \label{eq:UL}
\end{equation}
whereas the latter, which arises from the gas potential energy, is nonlinear: its specific expression depends on the nature of the fiber nonlinearity. 
It is important to note that $U_L/\mathcal{P}$ is bound by two extreme values: it is the sum of constant values (the $\beta_i$s) weighed by quantities whose sum is equal to 1 ($|c_i|^2/\mathcal{P}$).
The minimum value of $U_L/\mathcal{P}$ is $-\max\{\beta\} = - \beta_1$, and the minimum is $-\min\{\beta\} = - \beta_M $. 

\section*{Ideal photon gas}

In a weakly nonlinear regime, i.e., the regime where beam self-cleaning is observed, the contribution of $U_{NL}$ can be neglected, since $U_{NL}\ll U_L$. 
Under this approximation, thermal equilibrium states obey the Rayleigh-Jeans (RJ) law, i.e.,
\begin{equation}
    |c_i|^2 = - \frac{T}{\mu + \beta_i}.
    \label{eq:RJ}
\end{equation}
Note that, within this representation, $T$ and $\mu$ are expressed with the same unit of the propagation constants, and the power is considered to be a dimensionless quantity \cite{wu2019thermodynamic}. The values of $T$ and $\mu$ are found by combining eq. (\ref{eq:RJ}) with the equation of state
\begin{equation}
    U_L-\mu\mathcal{P} = MT.
    \label{eq:state}
\end{equation}
As demonstrated by Makris et al., there is only one physically meaningful solution for ($T,\mu$) \cite{makris2020statistical}. In addition, $T$ and $\mathcal{P}$ are linearly proportional, that is, varying the beam power does not affect the sign of the optical temperature (see Fig. \ref{fig:theo}b). Indeed, the latter is uniquely determined by $U_L$. Specifically, $T>0$ whenever $U_L<U_{L,c}$, with
\begin{equation}
    U_{L,c} = -\mathcal{P}\sum_{i=1}^M \beta_i/M.
\end{equation}
Viceversa, $U_L>U_{L,c}$ results in $T<0$. Finally, $U_L=U_{L,c}$ corresponds to the equipartition condition, i.e., $|c_i|^2=1/M$ $\forall i$, as illustrated in the configuration B of Fig. \ref{fig:toy}.

\section*{Real photon gas}

In a strongly nonlinear regime or dense photon gas, i.e., whenever $U_{NL}$ is not negligible, there is no closed form for the mode equilibrium distribution, which should replace the RJ law of the weakly nonlinear case. The equilibrium power fraction of any given mode depends on the power of virtually all other modes.
In short, even in the simplest case of a purely Kerr type of nonlinearity, finding the exact mode equilibrium distribution for a dense photon gas is quite challenging from the numerical point of view, and it is a daunting task to solve analytically.
Nevertheless, to explore the physics, one may rely on approximations, whose validity can be verified a posteriori by the experiment. In their description of the Joule–Thomson expansion, Kirsch et al. included the contribution to the potential energy by the Kerr nonlinearity, and considered the extreme cases where $U_{NL}$ takes either its maximum or its minimum value \cite{kirsch2025observation}.

Here, we make stronger assumptions. Let us consider the following phenomenological expression for $U_{NL}$
\begin{equation}
    U_{NL} = \frac{\gamma}{2} \frac{\mathcal{P}^2}{M},
    \label{eq:UNL}
\end{equation}
where $\gamma$ is a normalized nonlinear coefficient ($\gamma \propto n_2$, where $n_2$ is the nonlinear refractive index ). This expression provides the simplest law for the nonlinear dependence of the potential energy: it scales quadratically with the number of photons (in a Kerr-like fashion) and, at the same time, it is inversely proportional to the gas volume. In this way, $U_{NL}/\mathcal{P}$ is a function of the gas density $\mathcal{P}/M$.
In addition, we keep the same statistics as in the weakly nonlinear regime, i.e., we still rely on the RJ law (\ref{eq:RJ}) and the equation of state (\ref{eq:state}). This is justified by the fact that since both $\mathcal{P}$ and $M$ are constant, the expression (\ref{eq:UNL}) for $U_{NL}$ does not change the nature of the conservation laws. As such, in the presence of such a nonlinear term, the mode power distribution still obeys the RJ law.

The key idea of our approach consists of modifying the equation of state of the photon gas (\ref{eq:state}) as follows
\begin{equation}
    U_L+U_{NL} - \mu \mathcal{P} = MT.
    \label{eq:state_NL}
\end{equation}

We underline that the equation of state (\ref{eq:state_NL}) cannot be directly derived from the RJ law (\ref{eq:RJ}) combined with the conservation laws of $\mathcal{P}$ and $U$. Nonetheless, it has to be noted that the model does not violate the extensivity of the entropy. This can be understood by observing that, according to eq. (\ref{eq:state_NL}), the values of the thermodynamic parameters $T$ and $\mu$ do not change if the extensive variables $U_L$, $\mathcal{P}$, and $M$ are multiplied by a constant factor. Moreover, to ensure the consistency of the equation of state (\ref{eq:state_NL}) and the validity of the RJ distribution, $U_L$ should be intended as the internal energy at low powers, i.e., in the absence of nonlinearity ($\gamma = 0$).

\begin{figure}[!ht]
    \centering
    \includegraphics[width=0.7\linewidth]{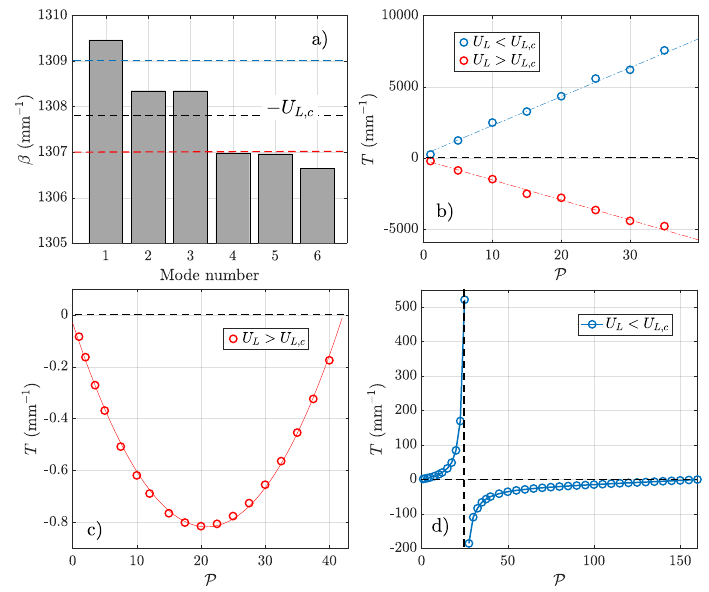}
 \caption{(a) Propagation constant vs. mode number. The black dashed horizontal line corresponds to the value of $-U_{L,c}$. (b-d) $T$ vs. $\mathcal{P}$ obtained solving Eq. (\ref{eq:RJ}) together with Eq. (\ref{eq:state_NL}) using a nonlinear solver (see Methods) in the weakly nonlinear regime (b) and strongly nonlinear regime (c,d). The dashed lines in (b) are linear fit. The solid lines in (c) and (d) are guides for the eye. 
 In (c) and (d), we used $\gamma = 0.05$. The values of $U_{L}$ used in (b-d) are indicated in (a) with a color code.
 }
    \label{fig:theo}
\end{figure}

In short, within this model, the presence of nonlinearity simply adds to $U_L$, leading to a nontrivial trend of $T$ vs. $\mathcal{P}$.
Let us take a look first at the case of negative absolute temperatures, i.e., $U_{L}>U_{L,c}$, which is shown in Fig. \ref{fig:theo}c. At low powers, we recover the linear decrease of $T$ with $\mathcal{P}$. However, when $\mathcal{P}$ grows larger, the curve eventually bends, it reaches a minimum, and eventually $T$ starts growing up to $T=0^-$. The zero minus temperature is reached at a critical value of power $\mathcal{P}$, which corresponds to the extreme condition $U_L/\mathcal{P} = - \beta_M$ at $U_{L}>U_{L,c}$ (i.e., only the highest-order mode is populated). Such an extreme condition sets the limit of validity of our model as at higher power there are no physical solutions for the thermodynamic parameters.

Things become even more interesting whenever $U_{L}<U_{L,c}$, i.e., when at low powers the temperature is positive (cfr. Fig. \ref{fig:theo}d). As the power grows larger, the temperature increases, up to diverging at a second critical value of power, corresponding to $U_L = U_{L,c}$, and then the sign of the temperature flips. Note that this divergence of $T$ is not due to the presence of any singularity in the model, since the Lagrange multiplier associated with the conservation of momentum is $1/T$. In addition, as $\mathcal{P}$ grows even larger, now $T$ tends to zero from negative values, i.e., again there is a power value for which the $M$-th mode is the only populated mode. This corresponds to the condition $U_L/\mathcal{P} = -\beta_1$ at $U_{L}<U_{L,c}$.

Interestingly, the results shown in Fig. \ref{fig:theo}c,d indicate that, above a certain power threshold, the highest-order mode is always predominant, regardless of the sign of $U_{L}-U_{L,c}$, i.e., irrespective of the injection conditions in the experiments! Such a negative absolute temperature attractor manifests at power levels which remain far below the critical power for self-focusing. As a matter of fact, $U_{NL}$ remains about two orders of magnitude lower than $U_L$. In essence, the striking evolution of $T$ vs. $\mathcal{P}$ results from their strongly nonlinear relationship, as it occurs even for an ideal photon gas \cite{wu2019thermodynamic,mangini2023high} in the presence of a small seed of nonlinearity.

\section*{Modal distribution inversion}

As previously discussed, the occurrence of the modal distribution inversion depends on the value of $\mathcal{P}/M$, i.e., on the photon gas density. One may increase the gas density either by enhancing the number of photons, i.e., by increasing the beam power, or by reducing the box volume, i.e., the number of modes. The former solution, however, is limited by the occurrence of the critical self-focusing, that leads to material damage. Therefore, in our experiments, we used a step-index silica optical fiber span that supports only a few guided modes. In this way, we could observe the modal distribution inversion at power levels well below the silica damage threshold.

\begin{figure}[!h]
    \centering
\includegraphics[width=1.0\linewidth]{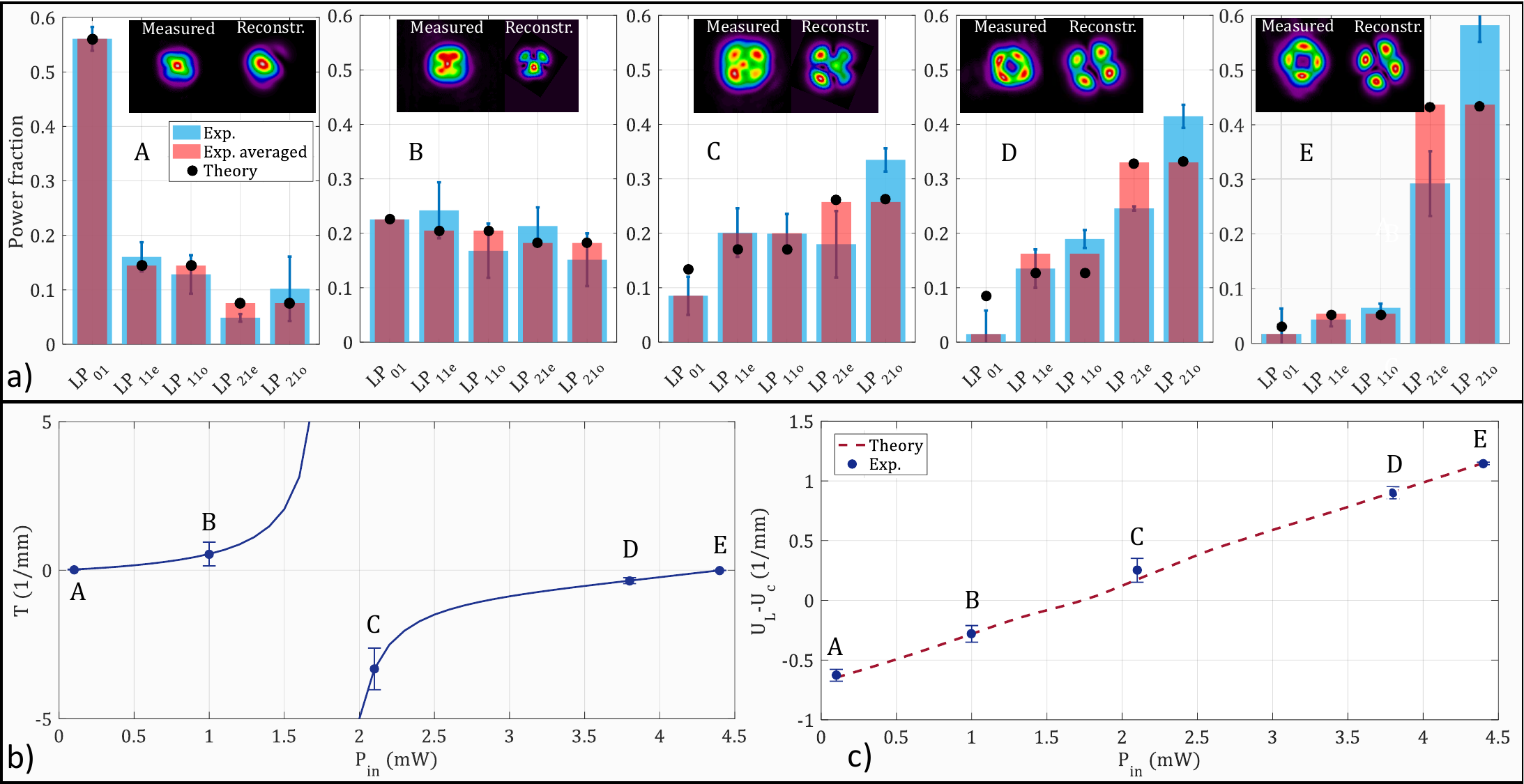}
    \caption{a) Modal distribution inversion in a 15 m long optical fiber. 
    The image pairs in the insets 
    correspond to the measured near field at the fiber output (left) and its reconstruction (right). b) Power and temperature values corresponding to states A-E are superimposed to the theoretical curve (blue line). 
    The value of $\gamma$ is 2.04. c) Comparison between experimental and theoretical values of the internal energy.
    }
    \label{fig:main_res}
\end{figure}

We used transform-limited light pulses with a tunable central wavelength from the signal of an optical parametric amplifier,
and a pulse duration of about 70 fs. At $\lambda$ = 700 nm, the fiber supports five modes per polarization, as shown in Fig. \ref{fig:toy}. In addition, the use of sub-100 fs pulses allowed us to minimize possible Raman contributions to the nonlinear beam dynamics. The laser beam, having a quasi-Gaussian shape, was injected into the center of the fiber core. In this way, owing to its radial symmetry, we ensured that the fundamental mode (LP${}_{01}$) was the most populated one at the fiber input. 
Using a holographic mode decomposition setup, we determined the mode occupancy distribution at the fiber output, for different input power values. The experimental results are shown as cyan bars in Fig. \ref{fig:main_res}a. The accuracy of the decomposition can be appreciated by comparing the measured beam profiles along with their reconstructions in the insets of Fig. \ref{fig:main_res}a. 

To make a fair comparison with theoretical predictions, we computed the average mode power fraction of degenerate modes, i.e., those modes that have the same propagation constant (see pink bars in Fig. \ref{fig:main_res}a). The theoretical values, indicated as full circles in Fig. \ref{fig:main_res}a, are calculated by combining Eq. (\ref{eq:RJ}) with the equation of state (\ref{eq:state_NL}). As can be seen, we found excellent agreement between experiments and theory. Indeed, we observed the predicted power-induced inversion of the modal distribution: at low powers, the fundamental mode predominates (state A); as the power grows larger, the modal distribution goes through an intermediate equipartition state (state B), until it reaches a state where the two highest order modes (the degenerate pair LP${}_{21e}$ and LP${}_{21o}$) have the largest occupancy (states C and D); at the highest power levels that we could launch without damaging the fiber, these modes hold about 90\% of the total power (state E).
The latter state corresponds to a spatial beam freezing, i.e., a state whose associated temperature is negative and close to zero. The evolution of the optical temperature vs. input power is illustrated in 
Fig. \ref{fig:main_res}b: here we display the continuous theoretical evolution as in Fig. \ref{fig:theo}d (dashed curves) along with the values corresponding to the five panels in Fig. \ref{fig:main_res}a (black dots). Whereas, in Fig. \ref{fig:main_res}c we compare the experimental and theoretical values of the internal energy: as can be seen, there is an excellent quantitative 
agreement.

It is also interesting to notice that the experimental mode power distribution follows the RJ law even at the lowest powers. This is because our fiber span was long enough to ensure that beam thermalization was reached in all cases. This aspect was further investigated with a cut-back experiment, as discussed in the Supplementary Note 1. In addition, we verified that the power flow from the fundamental to the highest-order modes do not result in light leakage into the cladding. Indeed, the output to input power ratio remains a constant across the whole modal distribution inversion (see Supplementary Note 2). Finally, we underline that even in the present case of dense photon gas
the nonlinear part of the internal energy $U_{NL}$ always remains much lower than $U_L$, in accordance with the theoretical description of the modal distribution inversion \cite{pourbeyram2022direct}.

\section*{Robustness of spatial beam freezing}

\begin{figure}[!h]
    \centering
\includegraphics[width=0.7\linewidth]{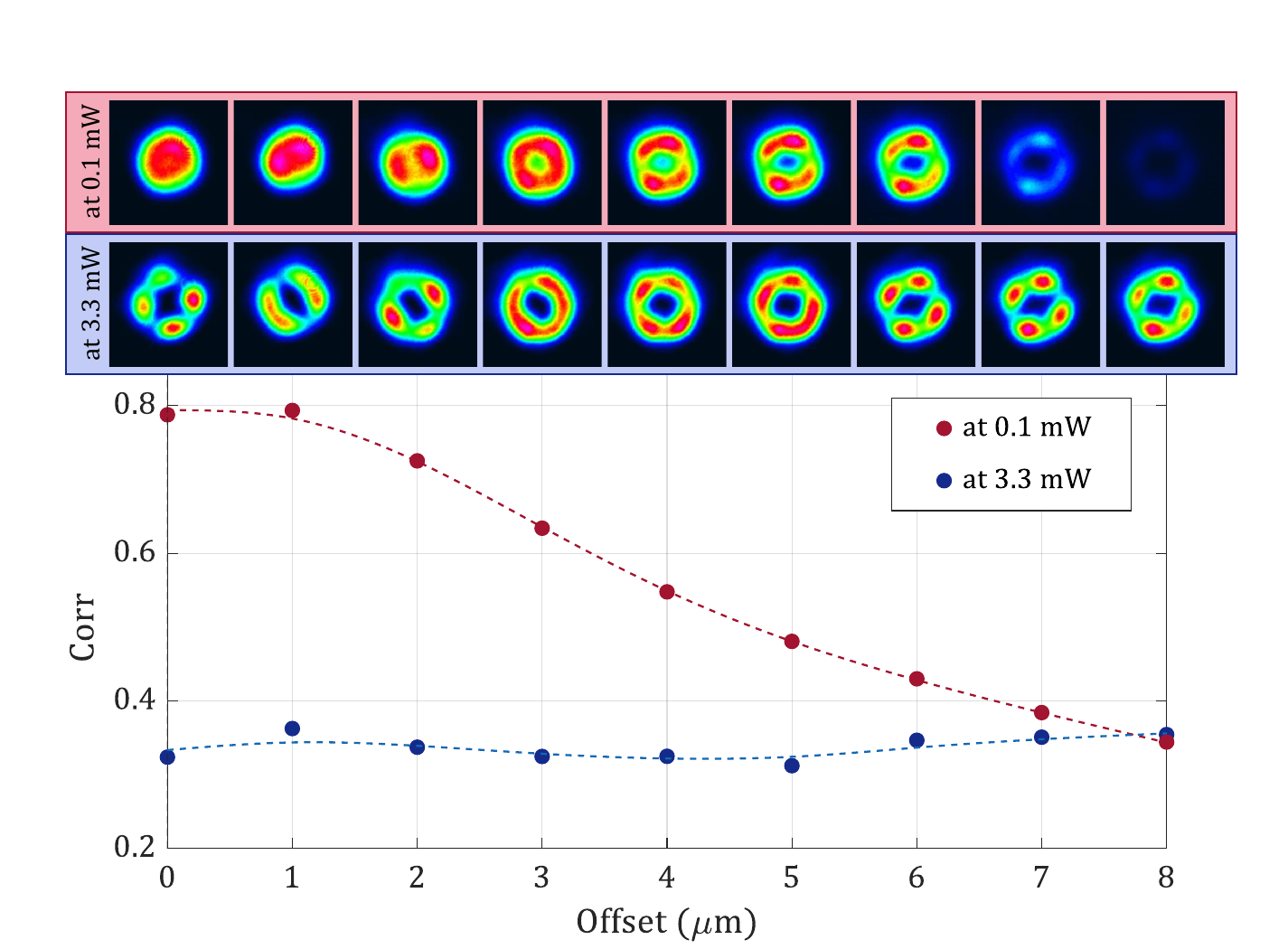}
    \caption{Robustness of spatial beam freezing in a 50 m long optical fiber. On the top: variation of the output near field vs. the offset between the laser beam direction and the fiber axis. On the bottom: Corresponding correlation with the fundamental mode calculated via eq. (\ref{eq:cor}).}
    \label{fig:offset}
\end{figure}

One of the most intriguing features of Van der Waals photon gases is the fact that, if the power is high enough, one always ends by getting spatial beam freezing at the fiber output. 
Indeed, no matter the sign of the temperature at low power, i.e., no matter the injection conditions of light into the fiber, eventually the optical temperature always approaches zero from the bottom (see Fig. \ref{fig:theo}c,d).
This was experimentally proven by introducing an offset between the fiber axis and the laser beam direction. As can be seen in Fig. \ref{fig:offset}, in a quasi-linear regime (red stripe) the output beam has a bell-shape in the absence of an offset. However, as the offset is introduced, the beam changes its shape: it acquires an LP${}_{11}$-like profile at an offset of 2 $\mu$m, and resembles the LP${}_{21}$ mode at larger offsets. 
Conversely, at sufficiently high powers (blue stripe in Fig. \ref{fig:offset}), the highest-order mode (LP${}_{21}$) is always the predominant one in the output beam. To provide a quantitative information, we computed the correlation of the measured near-field with the fundamental mode.
The result is shown at the bottom of Fig. \ref{fig:offset}. The correlation goes down from nearly 80\% to less than 40\% at low powers. In contrast, at high powers, the correlation remains below 40\%, and it is practically independent of the input beam offset, which is again in agreement with theoretical predictions.

This result indicates that $T=0^-$ is a sort of stable beam attractor at high powers. This provides spatial beam freezing with extreme robustness. In fact, in addition to the light injection conditions into the fiber, spatial beam freezing turns out to be virtually insensitive to external perturbations (such as fiber bending and squeezing, see the video in the Supplementary Materials).

\section*{Polychromatic freezing}

\begin{figure}[!h]
    \centering
\includegraphics[width=1.0\linewidth]{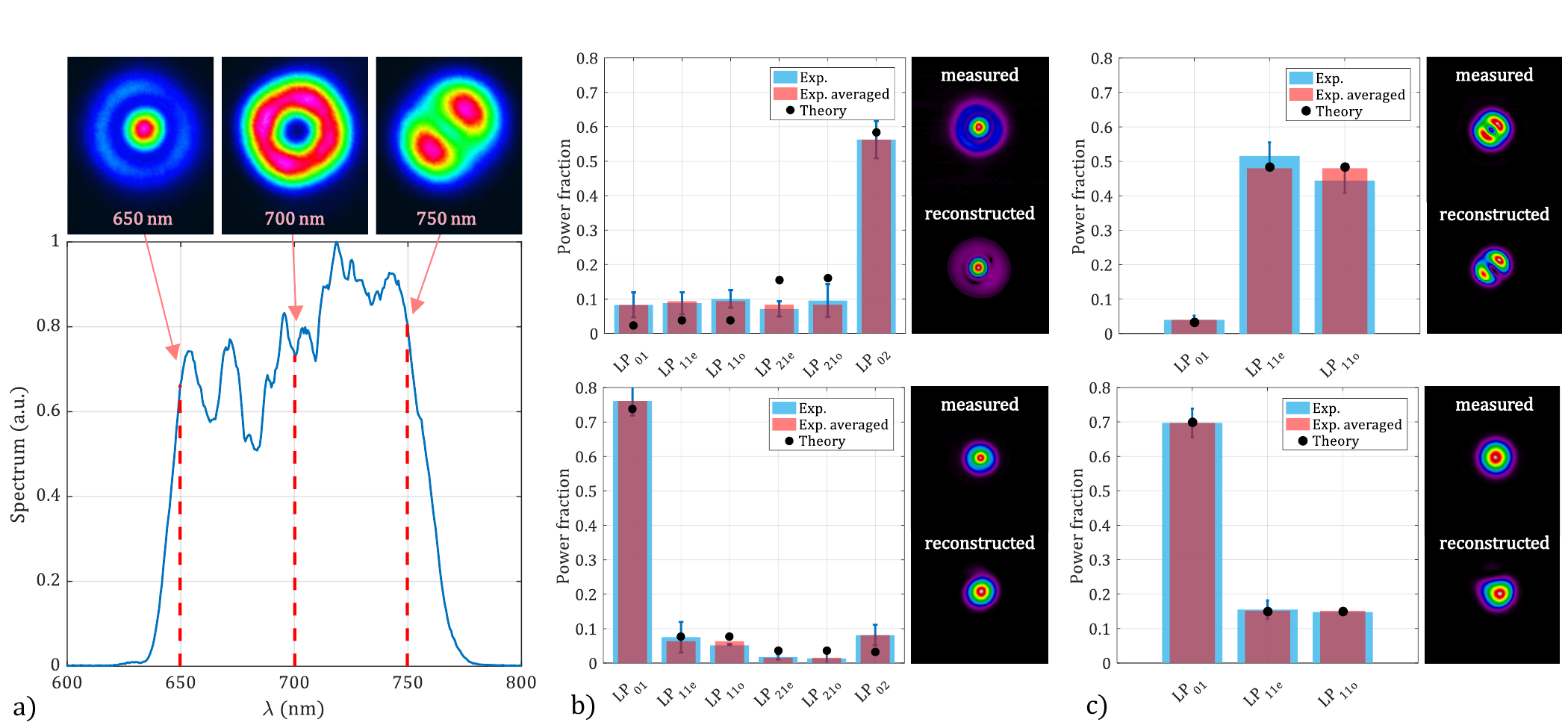}
    \caption{Spatial beam freezing with varying number of modes in a 50 m long optical fiber. (a) Output spectrum for 5~mW average power of the input beam with input wavelength $\lambda$ = 700 nm. The insets show the beam profile after filtering into the indicated spectral regions. (b) Mode distribution at low (bottom) and high power (top) and corresponding measured and reconstructed near-field profiles; here the input wavelength is $\lambda$ = 650 nm. (c) Same as (b) with input wavelength $\lambda$ = 750 nm.}
    \label{fig:wavelength}
\end{figure}

The validity of the thermodynamic approach to nonlinear beam propagation is intrinsically limited to monochromatic waves. However, when dealing with ultrashort pulses with peak power levels so high to permit modal distribution inversion, temporal effects are no longer negligible. In particular, since nonlinear beam dynamics is dominated by the Kerr effect, the evolution in the time domain corresponds to spectral broadening driven by self-phase modulation. This is shown in Fig. \ref{fig:wavelength}a. 
Strikingly, we observed that the thermodynamic approach remains valid across each slice of the entire frequency spectrum of the output beam. Namely, by passing the output spectrum through different bandpass filters, we always observed the occurrence of beam freezing, albeit at the highest-order mode that is guided at that particular wavelength (cfr. the insets in Fig. \ref{fig:wavelength}a)! For example, $M = 5$ modes are guided at $\lambda$ = 700 nm. Whereas, $M=3$ at $\lambda$ = 750 nm, and $M = 6$ at $\lambda$ = 650 nm.

Those observations are corroborated by varying the input laser carrier wavelength. In Fig. \ref{fig:wavelength}b and c, we show the mode power distribution inversion which occurs when using laser pulses at 650 and 750 nm, respectively. As can be seen, in both cases, the beam has a Gaussian-like shape at low powers, for the fundamental mode predominates. The other way around, at high powers, the beam approaches the LP${}_{11}$ and the LP${}_{02}$ modes at 750 or 650 nm, respectively.

These results indicate that, despite its monochromatic nature, still the thermodynamic approach works fine in describing all of our experimental observations. Similarly to the case of beam self-cleaning and wave thermalization combined with supercontinuum generation \cite{eslami2022two,Krupa:16}, here we found that spatial beam freezing spreads into the spectral domain.

In conclusion, we theoretically proposed and experimentally demonstrated that a Van der Waals photon gas in a multimode optical system eventually reaches negative and close to zero temperature if its density becomes too elevated. Incidentally, to the best of our knowledge, this is the first time that a statistical mechanics approach to beam propagation in nonlinear step-index fiber has been experimentally reported. 
In general, wave turbulence theory predicts that reaching thermal equilibrium in multimode optical fibers with a step-index profile is challenging to achieve in practice \cite{garnier2020wave}.
Finally, our findings enable the conversion of the control on the beam power into that of its spatial features. Regarding applications, spatial beam freezing can be exploited as a novel mechanism for spatial mode locking and will enable the development of spatially tunable high-power fiber lasers.



\section*{Methods}
\subsection*{Experiments}

We used a standard SMF-28 fiber from Alcatel that has a wavelength cutoff of around 1450 nm \cite{mangini2021experimental}. The fiber input and output ends were held on a three-axes stage that allowed micrometric shifts. As a pump source, we used a hybrid optical parametric amplifier (OPA) of a white-light continuum (Lightconversion Orpheus-F), pumped by a femtosecond Yb-based laser (Lightconversion Pharos-SP-HP) with a repetition rate of 100 kHz. The beam at the fiber input was linearly polarized. The bandpass filters used were from Thorlabs and have 10 nm full-width-half-maximum.
The beam intensity profile and power were measured with standard CCD cameras (Gentec Beamage M2) and power meters (fast photodiode by Thorlabs and thermopile by Gentec), respectively. 

Mode decomposition experiments were carried out using the holographic method described in Ref. \cite{gervaziev2020mode}. In each measurement, we placed a bandpass filter at the pump wavelength at the fiber output.
In all MD plots, we report the average value of the mode occupancy and its standard deviation when repeating the same reconstruction algorithm by moving the center of the output beam of $\pm$1 pixel in each direction with respect to the center position of the CCD camera of the MD device. This standard deviation was used to determine the error bars presented in the plots.
Finally, the spectra at the fiber output were collected using a miniature fiber optics spectrometer (Avaspec-2048L).

\subsection*{Determination of thermodynamic parameters}

The thermodynamic parameters were calculated by putting together the RJ distribution, eq. (\ref{eq:RJ}), and the equation of state in the strongly nonlinear regime, eq. (\ref{eq:state_NL}). This returns:
\begin{equation}
    \sum_{i=1}^M \frac{T/\mathcal{P}}{M(T/\mathcal{P})-U_{L}/\mathcal{P}-(\gamma\mathcal{P}/2M)-\beta_i}=1.
    \label{eq:TvsP_NL}
\end{equation}
In the absence of nonlinearity ($\gamma$=0), $T$ and $\mathcal{P}$ are linearly proportional, for only the ratio ($T/\mathcal{P}$) appears in the equation, and $U_L/\mathcal{P}$ is a constant. Moreover, the thermodynamic parameters ($T,\mu$) are uniquely determined by $(U_L,\mathcal{P})$. The presence of $\gamma$ breaks the linear relationship between $U_L$ and $\mathcal{P}$, and the pair ($T,\mu$) can no longer be determined by the mere knowledge of ($U_L,\mathcal{P}$). 

In practice, we determined $T$ from the experimental values of the mode occupancy as follows. At first, we computed $U_L$ by using eq.~(\ref{eq:UL}) from the data at the lowest power. Such a value of $U_L$ is kept as a constant for all power levels, when determining the thermodynamic parameters with eq.~(\ref{eq:TvsP_NL}). Specifically, 
we extracted the value of $\gamma$ via nonlinear fit of eq. (\ref{eq:TvsP_NL}) at the highest input power. At this point, we fixed the value of $\gamma$ and used eq. (\ref{eq:TvsP_NL}) to determine $T$ corresponding to each input power. Within this procedure, the values of the thermodynamic parameters are determined by just two sets of experimental data, i.e., those at the highest and the lowest power. Once $T$ had been known, the chemical potential was determined by the equation of state (\ref{eq:state_NL}), and the theoretical distribution was then calculated using the RJ law (\ref{eq:RJ}). 


\subsection*{Correlation with the fundamental mode}

We calculated the correlation between the intensity of the output beam ($I$) and that of the fundamental mode of the SMF-28 fiber ($I_0$), using the following formula:
\begin{equation}
    Cor = \frac{\int I(x,y) I_0(x,y) dx dy}{\sqrt{\int I(x,y)^2 dx dy\int I_0(x,y)^2 dx dy}}.
    \label{eq:cor}
\end{equation}
Both $I$ and $I_0$ were normalized in such a way that their integral over the CCD camera area is equal to 1.

\subsection*{Data availability}

The data that support the findings of this study are available from the corresponding author upon reasonable request.

\section*{Acknowledgments}

This work was supported by the European Innovation Council (101185664), the EU - Next Generation EU under the Italian National Recovery and Resilience Plan (NRRP), Mission 4, CUP B53C22004050001, partnership on “Telecommunications of the Future” (PE00000001 - program “RESTART”), the Italian Ministerial grant PRIN2022 "SAFE" (2022ESAC3K), the Deutsche Forschunggemeinschaft (DFG - 505515860 -Quadcomb), and the Russian Science Foundation (21-72-30024-$\Pi$).

We acknowledge Siria Boni, Alessandro Ciorra, Alexis Sparapani, Yifan Sun, and Georgios Pyrialakos for fruitful discussions.

\section*{Author contribution statement}

M.F. and F.M. initiated the idea. F.M. interpreted the experimental results within the thermodynamic framework. M.F. developed the theoretical model. F.M, M.F., K.S., and W.G. carried out experiments. F.M, M.F., S.W., and K.S performed the formal analysis of the data. V.C. provided optical fiber characterization. M.G., D.K., and S.B. provided the mode decomposition setup. S.W. supervised and administered the project. All the authors discussed the results and contributed to the writing of the manuscript. 

\section*{Competing interest}

The authors declare no competing interests.

\section*{Supplementary note 1: Cut-back experiment}
\vspace{-0.4cm}

\begin{figure}[!ht]
    \centering   \includegraphics[width=1\linewidth]{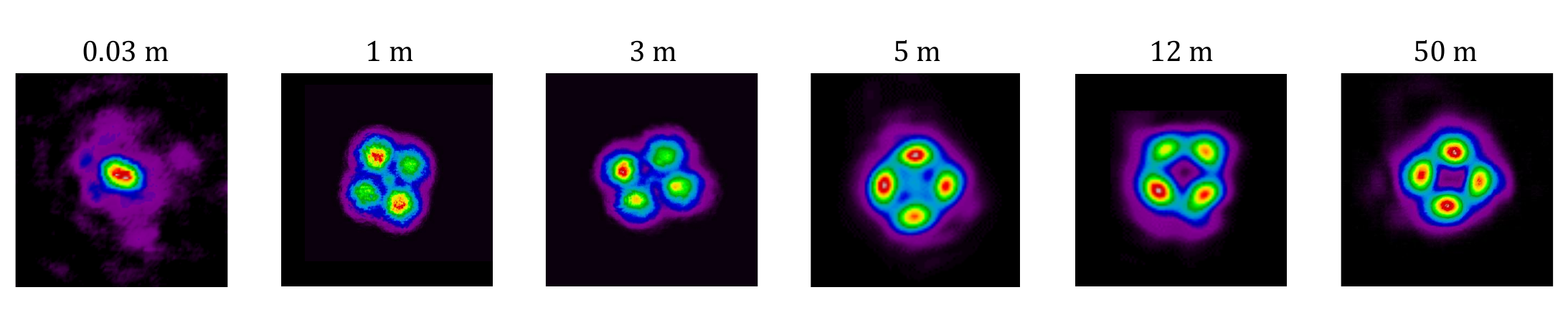}
    \caption{Example of beam evolution along the fiber, for an average input power of approximately 4 mW for all the cases. Output near-fields are measured from fiber segments of different lengths. The beam at the fiber input was injected straight onto the fiber axis, i.e., the fundamental mode is the most populated.}
    \label{fig:S1}
\end{figure}

To determine the range of fiber length which leads to beam thermalization, we performed a cutback experiment. In this way, we could analyze the evolution of output near-field as a function of input power, from fiber segments of different lengths. Our results indicate that a sufficiently long fiber is always required, in order to observe the flip of the temperature sign.
We found that thermalization occurs after propagation through fiber lengths on the order of of a few meters.
This is presented in Fig.~\ref{fig:S1}.
In experiments involving just a short (cm-long) fiber segment the beam always retained its initial input Gaussian profile, since it did not reach thermalization yet.
The beam only exhibits the profile of the highest-order mode LP$_{2,1}$ after at least one meter of propagation, and it maintains this shape thereafter.

\section*{ Supplementary note 2: Experimental evidence of power conservation}
\vspace{-0.4cm}

\begin{figure}[!ht]
    \centering   \includegraphics[width=0.4\linewidth]{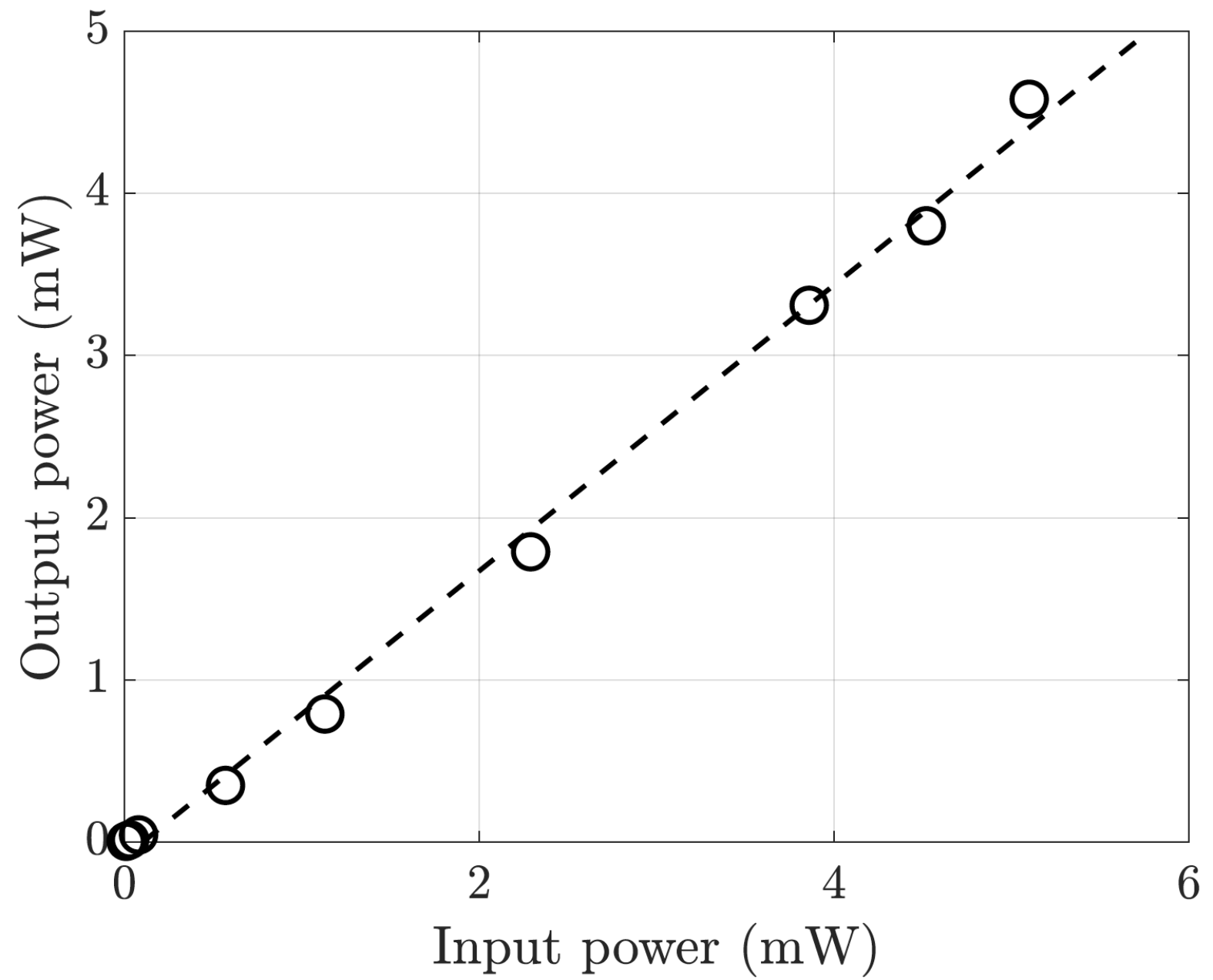}
    \caption{Experimental evidence of power conservation. Output vs. input power. The dashed line is a linear fit, with linear correlation coefficient above 99\%.}
    \label{fig:S2}
\end{figure}

To verify that there was no power leakage from high-order modes into the cladding at the occurrence of modal distribution inversion, we always monitored the output power vs. input power. The result is shown in Fig. \ref{fig:S2} in the case of a SMF-28 fiber about 50 m long, which was coiled in order to maximize the core-cladding mode coupling. As can be seen, a linear trend was observed, as highlighted by the dashed line. This indicates that light remained confined into the core for all of the power levels which were used in our experiments.

\bibliographystyle{apsrev4-1}
\bibliography{main}

\end{document}